\documentclass[epj]{svjour}
\usepackage{graphics}
\begin{document}
\title{Top Properties and Rare Decays from the Tevatron}
%\author{First author\inst{1} \and Second author\inst{2}% etc
\author{Arnulf Quadt\inst{1} \inst{2}
}                     % Do not remove
\institute{Physikalisches Institut, Universit\"at Bonn, 
           Nu{\ss}allee 12, D-53115 Bonn, Germany \and 
           University of Rochester, New York, c/o Fermilab - P.O. Box 500,
           60510, IL, USA}
\date{Received: date / Revised version: date}
\abstract{The top quark is the most recently discovered quark.
  Relatively little is known about its properties so far. Due to its
  very large mass of about $175\;\rm GeV/c^2$, the top quark behaves
  differently from all other quarks and provides a unique environment
  for tests of the Standard Model. Furthermore, it is believed to
  yield sensitivity to physics beyond the Standard Model. This report
  discusses the latest measurements and studies of top quark
  properties and rare decays from the Tevatron in Run~II.
\PACS{
      {PACS-key}{discribing text of that key}   \and
      {PACS-key}{discribing text of that key}
     } % end of PACS codes
} %end of abstract
\maketitle
\section{Introduction}
\label{intro}
The top quark discovery in 1995 by the experiments CDF and D\O\ 
\cite{tdiscovery} defines the start of the exciting era of top quark
physics at the Tevatron. After very successful upgrades of the
$p\bar{p}$ collider Tevatron for higher beam energy and luminosity and
of both experiments for faster readout and trigger electronics, better
tracking and muon detection, data taking in Run~II started in the year
2001. Since then, the Tevatron provided more than $1\;\rm fb^{-1}$ of
$p\bar{p}$ collision data at $\sqrt{s}=1.96\;\rm TeV$ to each
experiment. At present, up to $370\;\rm pb^{-1}$ have been analyzed in
top quark studies.

Top quark physics at the Tevatron can be divided into the following
categories: 1) top quark production, 2) fundamental properties of the
top quark, 3) top quark interactions to gauge bosons, 4) anomalous top
quark production, 5) anomalous top quark decays, and 6) new physics in
events with $t\bar{t}$ topology.

The first category, the top quark production, is studied via the
measurements of the strong $t\bar{t}$ production cross section and the
search for the electroweak single-top production, in the Standard
Model (SM) expected to be around $7\;\rm pb$ and $\approx 3\;\rm pb$,
respectively. Measurements of the $t\bar{t}$ production cross section
have been performed in many different top quark decay modes. The
results are found to be consistent between the two experiments, all
channels and with the Standard Model (SM) expectation within a
combined precision of $\approx 14\%$ \cite{talk_ebusato}.  The
corresponding data sets, quantitatively understood in terms of
selection efficiency and signal and background contribution form the
basis of all studies of properties and rare decays of the top quark.
Single-top production is expected to be observed with $1-2\;\rm
fb^{-1}$ of data \cite{talk_ataffard}.

The other categories are discussed in turn in this document in
Sections~\ref{sec:top_gauge_interaction} to
\ref{sec:new_physics_in_ttbar}. All limits are quoted at the 95\% CL
unless noted otherwise.

%Fundamental properties of the top quark include the measurement of the
%top quark mass \cite{talk_ttomura} and the measurement of the electric
%charge of the top quark (see Sect.~\ref{sec:top_charge}). 
%  
%The top quark interactions include studies of $t\bar{t}$ spin
%correlations (see Sect.~\ref{sec:tt_spincorr}), the measurement of
%$R=B(t\rightarrow Wb)/B(t\rightarrow Wq)$ (see
%Sect.~\ref{sec:top_bq_ratio}), studies of the $t\rightarrow \tau\nu q$
%decay (see Sect.~\ref{sec:top_tau_decay}), the measurement of the
%helicity of the $W$-boson in top quark decays (see
%Sect.~\ref{sec:top_wheli}), and the search for top quark decay via
%Flavor-Changing Neutral Current (FCNC) couplings (see
%Sect.~\ref{sec:top_fcnc}).
%      
%The anomalous top quark production includes the measurement of the
%cross section ratio $\sigma^{t\bar{t}}_{\ell
%  \ell}/\sigma^{t\bar{t}}_{\ell+jets}$ (see\linebreak
%Sect.~\ref{sec:top_xsec_ratio}), the search for anomalous kinematics
%in $t\bar{t}$ events (see Sect.~\ref{sec:top_kine}), and the search
%for $t\bar{t}$ production via narrow width, intermediate resonances
%(see Sect.~\ref{sec:top_ttreso}).
%       
%Anomalous top quark decays include the search for charged Higgs bosons
%in top quark decays.
%           
%And finally, new physics in events with $t\bar{t}$ topology includes
%the search for a fourth generation $t^\prime$ (see
%Sect.~\ref{sec:tprime}).

In the SM, assuming unitarity of the three-generation CKM matrix, the
matrix element $|V_{tb}|$ is found to be essentially unity. Therefore,
the top quark is expected to decay to a $W$-boson and a $b$-quark
nearly 100\% of the time. The $W$-boson subsequently decays either to
a pair of quarks or a lepton-neutrino pair. Depending on the lepton or
hadronic decay of the two $W$-bosons, the resulting event topologies
of $t\bar{t}$ decays are classified as all-jets channel (46.2\%),
lepton+jets ($\ell +$jets) channel (43.5\%), and dilepton ($\ell\ell$)
channel (10.3\%). Each decay topology contains at least two $b$-jets.
While $\ell$ in the above classification refers to $e$, $\mu$, or
$\tau$, most of the results to date rely on the $e$ and $\mu$
channels. Therefore, in what follows, $\ell$ will be used to refer to
$e$ or $\mu$, unless noted otherwise.

% ===============================================
\section{Top Quark Interactions to Gauge Bosons}
\label{sec:top_gauge_interaction}
% ===============================================

\subsection{Spin Correlation}\label{sec:tt_spincorr}
D\O\ has searched for evidence of spin correlation of $t\bar{t}$ pairs
\cite{d0_spincorr}. The $t$ and $\overline t$ are expected to be
unpolarized but to be correlated in their spins. Since top quarks
decay before hadronizing, their spins at production are transmitted to
their decay daughter particles. Spin correlation is studied by
analyzing the joint decay angular distribution of one $t$ daughter and
one $\overline t$ daughter. The sensitivity to top spin is greatest
when the daughters are down-type fermions (charged leptons or $d$-type
quarks), in which case, the joint distribution is
\begin{eqnarray}
{1\over \sigma}{d^2\sigma\over d(\cos \theta_+)d(\cos \theta_-)}&=&
{1+\kappa \cdot \cos \theta_+ \cdot \cos \theta_-\over 4},
\end{eqnarray}
where $\theta_+$ and $\theta_-$ are the angles of the daughters in the
top rest frames with respect to a particular spin quantization axis,
the optimal choice being the off-diagonal basis. In this basis, the SM
predicts maximum correlation with $\kappa = 0.88$ at the Tevatron. In
Run~I, D\O\ analyzed six dilepton events and obtained a likelihood as
a function of $\kappa$, which weakly favored the SM ($\kappa =0.88$)
over no correlation ($\kappa = 0$) or anti-correlation ($\kappa = -1$,
as would be expected for $t\overline t$ produced via an intermediate
scalar).  D\O\ quotes a limit $\kappa > -0.25$ at 68\% CL. With
improved statistics in the ongoing Run~II analyses, an observation of
$t\overline t$ spin correlation would support that the top quark
decays before hadronization and allow further test of the QCD
production mechanism.

\subsection{Measurement of \boldmath{$B(t\rightarrow Wb)/B(t\rightarrow Wq)$}}
\label{sec:top_bq_ratio}
\begin{figure}
\centerline{
\resizebox{0.44\textwidth}{!}{%
  \includegraphics{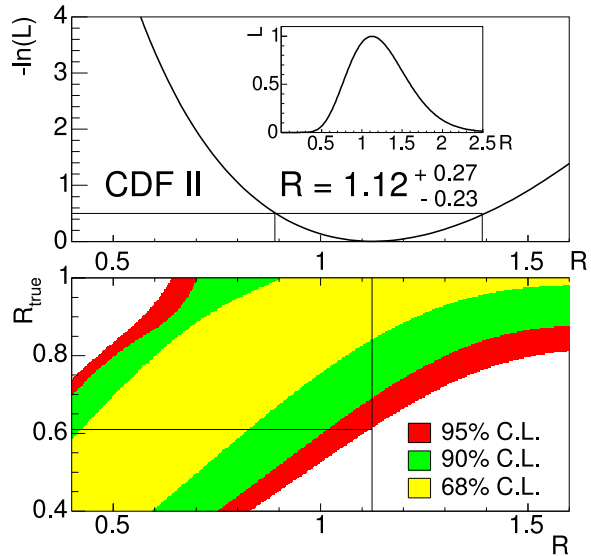}
}}
\caption{Top: CDF likelihood as a function of $R$ (inset) and its 
  negative logarithm. Bottom: Confidence level bands for $R_{true}$ as
  a function of $R$. The measurements of $R=1.12$ (vertical line)
  implies $R > 0.61$ (horizontal line).}
\label{fig:r_ratio}
\end{figure}

CDF and D\O\ report direct measurements of the $t\to Wb$ branching
ratio \cite{cdf_brratio,d0_brratio}. Comparing the number of events
with 0, 1 and 2 tagged $b$ jets in the lepton+jets channel, and for
CDF also in the dilepton channel, and using the known $b$-tagging
efficiency, the ratio $R=B(t\to Wb)/\sum_{q=d,s,b}B(t\to Wq)$ can be
extracted (Figure~\ref{fig:r_ratio}). D\O\ performs a simultaneous fit
for the production cross section $\sigma_{t\bar{t}}$ and the ratio
$R$. A deviation of $R$ from unity would imply either non-SM top
decay, a non-SM background to $t\bar{t}$ production, or a fourth
generation of quarks. Assuming that all top decays have a $W$ boson in
the final state, that only three generations of fermions exist, and
that the CKM matrix is unitary, CDF and D\O\ also extract the CKM
matrix-element $|V_{tb}|$. The results of these measurements are
summarized in Table~\ref{tab:br_ratio}. The top quark decay to $Wb$ is
indeed found to be dominant, although these studies are presently
limited by statistics and will profit from the upcoming larger data
sets.

\begin{table}
\caption{Measurements and lower limits of 
         $R = B(t \to Wb) / B(t \to Wq)$
         and $|V_{tb}|$ from CDF and D\O.}
\label{tab:br_ratio}
\centerline{
\begin{tabular}{llc}
\hline\noalign{\smallskip}
$R$ or $|V_{tb}|$ & Source & $\int\cal{L}$dt $(\rm pb^{-1})$   \\
\noalign{\smallskip}\hline\noalign{\smallskip}
$R = 1.12^{+0.27}_{-0.23}$&CDF~Run~\hfill II~\cite{cdf_brratio}&160\\
$R > 0.61$                &CDF~Run~\hfill II~\cite{cdf_brratio}&160\\
$R = 1.03^{+0.19}_{-0.17}$&D\O\phantom{F}~Run~II~\cite{d0_brratio}&230\\
$ R > 0.64$               &D\O\phantom{F}~Run~II~\cite{d0_brratio}&230\\
$|V_{tb}| > 0.78$         &CDF~Run~\hfill II~\cite{cdf_brratio}&160\\
$|V_{tb}| > 0.80$         &D\O\phantom{F}~Run~II~\cite{d0_brratio}&230\\
\noalign{\smallskip}\hline
\end{tabular}}
\end{table}

A more direct measurement of the $Wtb$ coupling constant will be
possible when enough data are accumulated to detect the $s$-channel
and $t$-channel single-top production processes \cite{talk_ataffard}.
The cross sections for these processes are proportional to $\vert
V_{tb} \vert^2$, and no assumption is needed on the number of families
or on the unitarity of the CKM matrix in extracting $\vert V_{tb}
\vert$.

\subsection{Study of \boldmath{$B(t\rightarrow \tau\nu q)$}}
\label{sec:top_tau_decay}
The SM's heavy third generation particles, the top and bottom quarks,
the tau and the tau neutrino are intriguing. The high energies
required to produce the third generation particles, particularly in
the case of the top quark, have resulted in the particles being the
least studied in the SM. Current measurements leave room for new
physics in the interactions and decays of these particles. The high
masses of the particles give rise to the hope that studying them could
help shed light on the origin of fermion masses. CDF measures the rate
of top-antitop events with a semi-leptonically decaying tau in
$t\bar{t} \rightarrow e \tau bb \nu \nu$ and $t\bar{t} \rightarrow \mu
\tau bb \nu \nu$ events in $200\;\rm pb^{-1}$ of Run~II data
\cite{cdf_top_tau}. Semi-leptonic tau decays account for 64\% of all
tau decays.  This analysis does not include taus decaying to electrons
or muons because their leptonic tau decays are difficult to
differentiate from prompt leptons. CDF compares the observed with the
predicted rate as a test of the SM. Many extensions to the SM predict
identical final states which could lead to an anomalous rate. For
example the charged Higgs decay from $t\bar{t}$, $t\bar{t} \rightarrow
H^\pm Wb\bar{b}$, $H^\pm \rightarrow \tau^\pm \nu_\tau$.  This
analysis is a search for any such anomalous processes that could show
up in the final state as an enhanced (or suppressed) rate for tau
leptons in top decays. The ratio $r_\tau \equiv B(t \to b \tau \nu) /
B_{SM}(t \to b \tau \nu)$ is found to be $r_\tau < 5.0$ and therefore
consistent with the SM.

\subsection{Measurement of the Helicity of the \boldmath{$W$}-Boson 
  in Top Quark Decays}\label{sec:top_wheli} 
\begin{figure}
\centerline{
\resizebox{0.44\textwidth}{!}{%
  \includegraphics{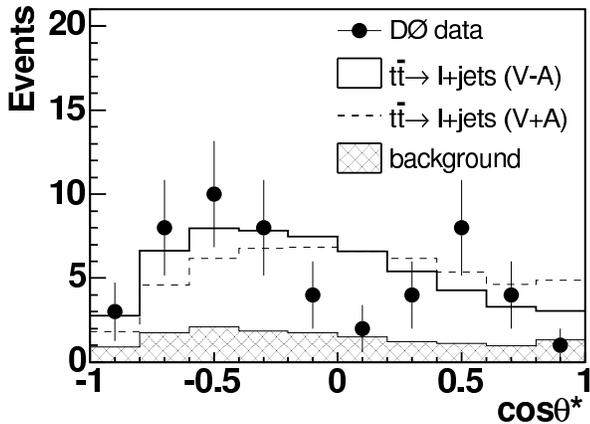}
}}
\caption{$\cos\theta^*$ distribution observed in the D\O\ 
  data along with the SM prediction (solid line) and a model with a
  pure $V$+$A$ interaction (dashed line) for the $b$-tagged
  lepton+jets sample.}
\label{fig:top_wheli}
\end{figure}

Studies of decay angular
distributions provide a direct check of the $V\hbox{--}A$ nature of
the $Wtb$ coupling and information on the relative coupling of
longitudinal and transverse $W$~bosons to the top quark. In the SM,
the fraction of decays to longitudinally polarized $W$ bosons is
expected to be ${\cal F}^{\rm{SM}}_0=x/(1+x)$, $x=m_t^2/2M_W^2$
(${\cal F}^{\rm{SM}}_0 \sim 70\%$ for $m_t=175$~GeV/$c^2$). Fractions
of left- or right-handed $W$ bosons are denoted as $\cal{F}_-$ and
$\cal{F}_+$, respectively. In the SM, $\cal{F}_-$ is expected to be
$\approx 30\%$ and ${\cal{F}_+} \approx 0\%$. CDF and D\O\ use various
techniques to measure the helicity of the $W$ boson in top quark
decays in lepton+jets events.
The first method uses a kinematic fit, similar to that used in the
lepton+jets mass analyses \cite{talk_ttomura}, but with the top quark
mass constrained to $175\;\rm GeV/c^2$, to improve the reconstruction
of final state observables and choose the assignment to quarks and
leptons as that with the lowest $\chi^2$. The distribution of the
helicity angle ($\cos \theta^*$) between the lepton and the $b$ quark
in the $W$ rest frame, provides the most direct measure of the $W$
helicity (Figure~\ref{fig:top_wheli}).
The second method ($p_T^\ell$) uses the different lepton $p_T$ spectra
from longitudinally or transversely polarized $W$-decays to determine
the relative contributions. This method is also used by both
experiments in the dilepton channel. 
A third method uses the invariant mass of the lepton and the $b$-quark
in top decays ($M_{\ell b}^2$) as an observable, which is directly
related to $\cos \theta^*$. Finally, the Matrix Element method (ME),
initially developed for the top quark mass measurement, has also been
used, forming a 2-dimensional likelihood ${\cal{L}}(m_{top}, {\cal
  F}_0)$, where the mass-dependence is integrated out so that only the
sensitivity to the $W$-helicity in the top quark decay is exploited.
The results of all CDF and D\O\ analyses, summarized in
Table~\ref{tab:top_wheli}, are in agreement with the SM expectation,
but within large statistical uncertainties.

\begin{table}
\caption{Measurement and upper limits of the $W$ helicity
         in top quark decays from CDF and D\O. The integrated luminosity 
         $\int\cal{L}$dt is given in units of $(\rm pb^{-1})$.}
\label{tab:top_wheli}
\centerline{
\begin{tabular}{lllc}
\hline\noalign{\smallskip}
$W$ helicity & Source & $\int\cal{L}$dt  & Method \\
\noalign{\smallskip}\hline\noalign{\smallskip}
${\cal F}_0 = 0.91 \pm 0.39$ &CDF~Run \phantom{I}I~\cite{cdf1_wheli}&106& $p_T^\ell$\\
${\cal F}_0 = 0.56 \pm 0.32$ &D\O\phantom{F}~Run \phantom{I}I~\cite{d01_wheli}&125 & ME\\
${\cal F}_0 = 0.74^{+0.22}_{-0.34}$ & CDF~Run II~\cite{cdf2_wheli}&200 & $M_{\ell b}^2+$$p_T^\ell$ \\ 
\noalign{\smallskip}\hline\noalign{\smallskip}
${\cal F}_+ <0.18$ &CDF~Run \phantom{I}I~\cite{cdf1.1_wheli}&110 &$M_{\ell b}^2$+$p_T^\ell$\\
${\cal F}_+ <0.27$&CDF~Run II~\cite{cdf2_wheli}&200 & $M_{\ell b}^2+$$p_T^\ell$\\
${\cal F}_+ <0.25$ &D\O\phantom{F}~Run II~\cite{d02_wheli} &230-370&$\cos\theta^*$$+$$p_T^\ell$\\
\noalign{\smallskip}\hline
\end{tabular}}
\end{table}

\subsection{Search for Top Quark Decay via FCNC Couplings}
\label{sec:top_fcnc}
Physics beyond the SM can manifest itself by altering the expected
rate of flavor-changing neutral-current (FCNC) interactions.  FCNC
decays of the top quark are of particular interest. The large mass of
the top quark suggests a strong connection with the electroweak
symmetry breaking sector. Evidence for unusual decays of the top quark
might provide insights into that mechanism. For the top quark, the
FCNC decays $t\rightarrow qZ$ and $t\rightarrow q\gamma$ (where $q$
denotes either a $c$- or a $u$-quark) are expected to be exceedingly
rare (branching fractions of $10^{-10}$ or smaller), since they are
suppressed by the GIM mechanism and any observation of these decays in
the available data sample would indicate new physics. In general, FCNC
interactions are present in models which contain an extended Higgs
sector, Supersymmety, dynamical breaking of the electroweak symmetry,
or an additional symmetry. 

CDF reported a search for flavor changing neutral current (FCNC)
decays of the top quark $t \to q \gamma$ and $t \to q Z$ in the Run~I
data \cite{cdf1_fcnc}. CDF assumes that one top decays via FCNC while
the other decays via $W b$. For the $t \to q \gamma$ search, two
signatures are examined, depending on whether the $W$ decays
leptonically or hadronically.  For leptonic $W$ decay, the signature
is $\gamma \ell$ and missing $E_T$ and two or more jets, while for
hadronic $W$ decay, it is $\gamma + \ge 4$ jets. In either case, one
of the jets must have a secondary vertex b tag. One event is observed
($\mu \gamma$) with an expected background of less than half an event,
giving an upper limit on the top branching ratio of $ B(t \to q
\gamma) < 3.2\%$. In the search for $t \to q Z$, CDF considers $Z \to
\mu \mu$ or $e e$ and $W \to q q^\prime$, giving a $Z$ + four jets
signature.  One $\mu \mu$ event is observed with an expected
background of 1.2 events, giving an upper limit on the top branching
ratio of $ B(t \to q Z) < 0.33$. These limits on top quark decay
branching ratios can be translated into limits on the flavor-changing
neutral current couplings $\kappa_{\gamma} < 0.42$ and $\kappa_Z <
0.73$. With $2\;\rm fb^{-1}$, CDF and D\O\ are expected to improve
their sensitivity to $\kappa_\gamma$ and to $\kappa_Z$ significantly
with the increased Run~II data set.

%Constraints on FCNC couplings of the top quark can also be obtained
%from searches for anomalous single-top production in $e^+e^-$
%collisions at LEP, via the process \linebreak $e^+e^- \rightarrow
%\gamma,Z^* \rightarrow t \overline q$ and its charge-conjugate
%($q=u,c$), or in $e^\pm p$ collisions at HERA, via the process $e^\pm
%u \rightarrow e^\pm t$.

% ===============================================
\section{Fundamental Properties of the Top Quark}
\label{sec:top_fundamental_properties}
% ===============================================

\subsection{Top Quark Mass}\label{sec:top_mass}
The Tevatron Electroweak Working Group has recently combined all
available direct measurements of the top quark mass yielding a new
world average of $m_{top} = 172.7\pm 2.9\;\rm GeV/c^2$
\cite{talk_ttomura,mtop_worldave}. The ultimate precision from the
Tevatron on the top mass measurement is expected to be better than
$2.0\;\rm GeV/c^2$ per experiment.

\subsection{Electric Charge of the Top Quark}\label{sec:top_charge}
The top quark is the only quark whose electric charge has not been
measured through a production threshold in $e^+e^-$ collisions.  Since
the CDF and D\O\ analyses on top quark production do not associate the
$b$, $\bar{b}$ and $W^\pm$ uniquely to the top or antitop, decays such
as $t \to W^+\bar{b}$, $\bar{t} \to W^-b$ are certainly conceivable. A
charge $4/3$ quark of this kind would be consistent with current
electroweak precision data. The $Z \to \ell^+ \ell^-$ and $Z \to
b\bar{b}$ data can be fitted with a top quark of mass $m_t = 270\;\rm
GeV/c^2$, provided that the right-handed $b$ quark mixes with the
isospin $+1/2$ component of an exotic doublet of charge $-1/3$ and
$-4/3$ quarks, $(Q_1, Q_4)_R$. CDF and D\O\ study the top quark charge
in double-tagged lepton+jets events. Assuming the top and antitop
quarks have equal but opposite electric charge, then reconstructing
the charge of the $b$-quark through jet charge discrimination
techniques, the $|Q_{top}| = 4/3$ and $|Q_{top}| = 2/3$ scenarios can
be differentiated. CDF and D\O\, both have already collected
sufficient data to obtain sensitivity to the $|Q_{top}| = 4/3$ case.
The analyses are ongoing, results are expected to be made public soon.

% ===============================================
\section{Anomalous Top Quark Production}
\label{sec:top_anomalous_prod}
% ===============================================

\subsection{Cross Section Ratio $\sigma_{\ell \ell} / \sigma_{\ell+jets}$}
\label{sec:top_xsec_ratio}
It is a priori not obvious, that the `top quark', observed in the
dilepton decay mode is identical to the `top quark' in the lepton+jets
decay mode. If both decay modes result exclusively from the decay of
the SM top quark, they should have the same production cross section.
If the production or the decay of the top quarks had non-SM
contributions, one mode might be enhanced with respect to the other.

CDF has measured the cross section ratio \linebreak $R_\sigma =
\sigma_{ll}/\sigma_{l+jets}$ of the $t\bar{t}$ production cross
section in the dilepton and the lepton+jets channels in $125\;\rm
pb^{-1}$ of Run~II data. CDF finds $R_\sigma = 1.45^{+0.83}_{-0.55}$
and $R_\sigma > 0.46\ (< 4.45)$, consistent with the SM.  This result
is also translated into generic top decay branching ratio limits. The
considered cases are a fully hadronic decay $t \rightarrow Xb$, where
$Br(X \rightarrow qq) = 100\%$ or a fully leptonic decay, i.e. $t
\rightarrow Yb$, where $Br(Y \rightarrow qq) = 100\%$. The limits on
$R_\sigma$ translate into limits on the fully hadronic or the fully
leptonic decay of the top quark as $Br(t \rightarrow Xb) < 0.46$ and
$Br(t \rightarrow Yb) < 0.47$.

\subsection{Anomalous Kinematics in $t\bar{t}$ Events}\label{sec:top_kine}
CDF reports a search for anomalous kinematics of $t\bar{t}$ dilepton
events in $193\;\rm pb^{-1}$ \cite{cdf2_ttkine}. A new {\it a priori}
technique has been developed, designed to isolate the subset of events
in a data sample which reveals the largest deviation from SM
expectation and to quantify the significance of this departure. Four
variables are considered: the missing transverse energy,
$\not\!\!E_T$, the transverse momentum of the leading lepton
$p_T^\ell$, the angle $\phi_{\ell m}$ between the leading lepton and
the direction of $\not\!\!E_T$ in the plane transverse to the beam,
and a variable $T$, representing how well the kinematics of an event
satisfy the $t\bar{t}$ decay hypothesis based on the expected and
observed $\not\!\!E_T$ vector. This method is especially sensitive to
data subsets that preferentially populate regions where new high-$p_T$
physics can be expected. No such subset is found.  Although the lepton
$p_T$ distribution exhibits a mild excess at low $p_T$, CDF determines
the level of consistency of the $t\bar{t}$ dilepton sample with the SM
expectation and finds a $p$-value of $1.0 - 4.5\%$, showing good
agreement with the SM.

This type of search for anomalous kinematics is presently statistics
limited and will improve with larger data sets.

\subsection{Top Production via Intermediate Resonances}
\label{sec:top_ttreso}
Motivated by the large mass of the top quark, several models suggest
that the top quark plays a role in the dynamics of electroweak
symmetry breaking. One example is topcolor, where a large top quark
mass can be generated through the formation of a dynamic $t\bar{t}$
condensate, $X$, which is formed by a new strong gauge force coupling
preferentially to the third generation. Another example is
topcolor-assisted technicolor, predicting a heavy $Z^\prime$ boson
that couples preferentially to the third generation of quarks with
cross sections expected to be visible at the Tevatron.  CDF and D\O\ 
have searched for $t\bar{t}$ production via intermediate,
narrow-width, heavy vector bosons $X$ in the lepton+jets channels.
The $t$ and $\bar{t}$ final states are identified through a kinematic
fit.  The possible $t\bar{t}$ production via an intermediate resonance
$X$ is sought for as a peak in the spectrum of the invariant
$t\bar{t}$ mass. CDF and D\O\ exclude narrow width heavy vector bosons
$X$ \cite{hill99} with mass $M_X < 480\;\rm GeV/c^2$ and $M_X <
560\;\rm GeV/c^2$, respectively, in Run~I \cite{cdfd01_ttreso}, and
$M_X < 680\;\rm GeV/c^2$ in D\O\ Run~II \cite{d0_4880}.

% ===============================================
\section{Anomalous Top Quark Decays}
\label{sec:top_anomalous_decay}
% ===============================================

\subsection{Search for Charged Higgs Boson in $t\bar{t}$ Decays}
\label{sec:t_hpmdecay}
Both CDF and D\O\ have searched for non-SM top decays, particularly
those expected in supersymmetric models, such as $t\to H^+ b$,
followed by $H^+\rightarrow \tau^+ \bar{\nu}$ or $c\overline s$. The
$t\to H^+ b$ branching ratio has a minimum at $\tan \beta =
\sqrt{m_t/m_b} \simeq 6$, and is large in the region of either $\tan
\beta\ll 6$ or $\tan \beta\gg 6$. In the former range, $H^+\to
c\overline s$ is dominant, while $H^+\to \tau^+ \bar{\nu}$ dominates
in the latter range. These studies are based either on direct searches
for these final states, or on top ``disappearance''. In the standard
lepton+jets or dilepton cross section analyses, any charged Higgs
decays are not detected as efficiently as $t\to W^\pm b$, primarily
because the selection criteria are optimized for the standard decays,
and because of the absence of energetic isolated leptons in Higgs
decays. A significant $t\to H^+ b$ contribution would give rise to
measured $t\bar{t}$ cross sections lower than the SM prediction
(assuming that non-SM contributions to $t\overline t$ production are
negligible).

\begin{figure}
\centerline{
\resizebox{0.44\textwidth}{!}{%
  \includegraphics{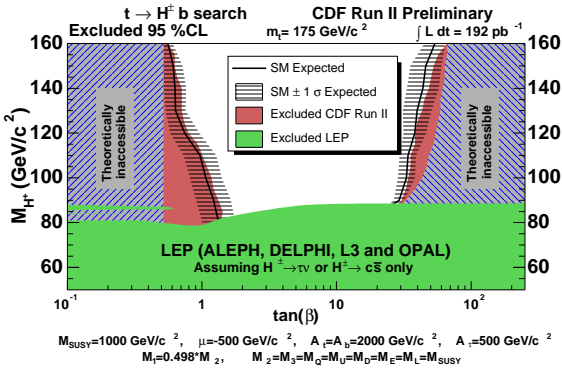}
}}
\caption{CDF exclusion region (red solid region) along with the
  expected exclusion limits (black solid line) and the 1-sigma
  confidence band around it in the ($M_{H^\pm}, \tan \beta$) plane.}
\label{fig:top_hpm}
\end{figure}

In Run~II, CDF has searched for charged Higgs production in dilepton,
lepton+jets and lepton+hadronic tau final states, considering possible
$H^+$ decays to $c\bar{s}$, $\tau\bar{\nu}$, $t^*b$ or $W^+h^0$ in
addition to the SM decay $t\rightarrow W^+b$ \cite{cdf2_hpm}.
Depending on the top and Higgs decay branching ratios, which are
scanned in a particular 2-Higgs Doublet benchmark Model, the number of
expected events in these decay channels can show an excess or deficit
when compared to SM expectations. A model-independent interpretation,
yields a limit of $B(t \to H^\pm b) < 0.91$ for $80\;{\rm GeV} <
m_{H^\pm} < 160\;\rm GeV$.  Stronger limits are set assuming specific
$H^+$ decay scenarios (see Figure~\ref{fig:top_hpm}).

% ===============================================
\section{New Physics in Events with $t\bar{t}$ Topology}
\label{sec:new_physics_in_ttbar}
% ===============================================

\subsection{Search for a Fourth Generation $t^\prime$ Quark}\label{sec:tprime}
Recent theoretical developments, such a Little Higgs Models, 2-Higgs
Doublet scenarios, $N=2$ SUSY models, or the ``beautiful mirror''
model \cite{beauty_mirror}, hypothesize the existence of a heavy
$t^\prime$. Assuming that such a new heavy $t^\prime$ quark is
pair-produced strongly, has mass greater than the top quark, and
decays promptly to $Wq$ final states, the final state event topology
is very similar to that of $t\bar{t}$ events, except that the
distribution of the total transverse energy $H_T$ would tend to larger
values.

CDF has performed a search for such a heavy $t^\prime$ quark in the
lepton+jets channel using $200\;\rm pb^{-1}$ of Run~II data
\cite{cdf2_tprime}. The observed $H_T$ distribution is compared to a
combination of SM background and $t\bar{t}$ signal, the latter with
floating normalization, plus a possible $t^\prime \bar{t}^\prime$
signal using a maximum likelihood fit, allowing to set upper cross
section limits for $t^\prime$ production as a function of the
$t^\prime$ mass. In comparison to the expected QCD $t^\prime
\bar{t}^\prime$ production cross section, these results are translated
into $t^\prime$ mass limits, ruling out a $t^\prime$ with mass greater
than about $175\;\rm GeV/c^2$, if the true top mass is about the same
value. For a smaller top mass the excluded $t^\prime$ mass is lower,
and vice versa for higher masses. The CDF limit on the $t^\prime$
production will steadily improve with more data in Run~II.

\section{Summary}
\label{sec:summary}
After the top quark discovery in Run~I and the re-esta\-blish\-ment of
the top quark signal with the upgraded detectors and improved analysis
techniques in the early Run~II, top quark physics at the Tevatron has
now entered the stage of detailed studies of the top quark properties.
A wealth of results on top quark properties in the SM as well as
searches for new top quark couplings and decays are becoming
available. This development is expected to even accelerate with $\ge
1\;\rm fb^{-1}$ of data being available to both, CDF and D\O, very
soon.

\section*{Acknowledgment}\label{sec:acknowledgement}
I thank to organizers of HCP2005 for a stimulating conference and
acknowledge the support by the Alexander von Humboldt Foundation and
the University of Rochester.

%
%% For one-column wide figures use
%\begin{figure}
%% Use the relevant command for your figure-insertion program
%% to insert the figure file.
%% For example, with the option graphics use
%\resizebox{0.75\textwidth}{!}{%
%  \includegraphics{leer.eps}
%}
%% If not, use
%%\vspace{5cm}       % Give the correct figure height in cm
%\end{figure}
%%
%% For two-column wide figures use
%\begin{figure*}
%% Use the relevant command for your figure-insertion program
%% to insert the figure file. See example above.
%% If not, use
%\vspace*{5cm}       % Give the correct figure height in cm
%\end{figure*}
%\begin{table}
%\begin{tabular}{lll}
%\hline\noalign{\smallskip}
%\end{tabular}
%% Or use
%\vspace*{5cm}  % with the correct table height
%\end{table}
%
% BibTeX users please use
% \bibliographystyle{}
% \bibliography{}

\end{document}